\documentstyle[12pt]{article}
\begin{document}
\newcommand{\be}{\begin{equation}}
\newcommand{\bi}{\bibitem}
\newcommand{\al}{\alpha}
\newcommand{\ea}{\alpha_{el}}
\newcommand{\as}{\alpha_s}
\newcommand{\bb}{\beta}
\newcommand{\bef}{\beta -{\rm function}}
\newcommand{\La}{\Lambda_{QCD}}
\newcommand{\ee}{\end{equation}}
\newcommand{\aQ}{\alpha_s(Q^2)}
\newcommand{\Nc}{N_{crit}}
\newcommand{\De}{\Delta}
\newcommand{\vp}{\varphi}
\newcommand{\la}{\lambda}
\newcommand{\bea}{\begin{eqnarray}}
\newcommand{\eea}{\end{eqnarray}}
\newcommand{\LQ}{\Lambda_{QCD}}
\setlength{\baselineskip}{0.30in}

\title{Notes on chiral hydrodynamics within effective theory approach
}
\author{  A.V. Sadofyev, V.I. Shevchenko, V.I. Zakharov\\
Institute of Theoretical and Experimental Physics, Moscow}

\maketitle

\begin{center}{\bf Abstract}\end{center}
   We address the issue of evaluating chiral effects (such as
   the newly discovered chiral separation) in hydrodynamic approximation.
   The main tool we use is effective theory
   which defines interaction in terms of chemical potentials $\mu,\mu_5$. In the lowest order
   in $\mu,\mu_5$ we reproduce recent results based on  thermodynamic considerations.
   In higher orders the results depend on details of infrared cutoff.
   Another point of   our  interest is an alternative way of the anomaly matching
   through introduction of effective scalar fields arising in the
   hydrodynamic approximation.

\section{Introduction.}

There is a clear renewed interest in the relativistic hydrodynamics, in view of its success in
describing properties of the (strongly interacting) quark-gluon plasma (see, e.g., \cite{teaney} and
references therein). Recently some new hydrodynamic phenomena resulted from existence of novel
transport coefficients have been found \cite{erdmenger,surowka}. One considers a medium made of
chiral  fermions such that there exist two conserved charges $Q$ and $Q_5$ and one can introduce
two
 corresponding chemical potentials: $\mu$ and $\mu_{5}$. Then in the standard hydrodynamic
approximation one would use for the phenomenological vector and axial-vector currents the following
textbook expressions:
\begin{equation}
J^{\mu}~=~ n\cdot u^{\mu}, ~~J_{5}^{\mu}~=~ n_5\cdot u^{\mu}~, \label{currents}
\end{equation}
where $n$ and $n_5$ are densities of
 particles with the corresponding charges,
while $u^{\mu}$ is the 4-velocity of an element of the liquid. As is argued in
\cite{erdmenger,surowka} one can add another term to the axial-vector current
\begin{equation}\label{effect}
\delta J_5^{\mu }~=~c_{\omega}\cdot \omega^{\mu}~,
\end{equation}
where $\omega^{\mu}\equiv \frac{1}{2}\epsilon^{\mu\nu\rho\sigma}u_{\nu}\partial_{\rho}u_{\sigma}$ and the new
term describes chirality separation through rotation. Indeed, if the liquid is rotated with angular
velocity ${\bf \Omega}$ then according to (\ref{effect}) there arises chiral current along the
vector ${\bf \Omega}$. A remarkable feature of the new term is that, according to \cite{surowka} the
coefficient $c_{\omega}$ (in our approach see (\ref{anomaly1})) is uniquely fixed by the requirement of positivity of the divergence of the
entropy current, $\partial_{\mu}s^{\mu}>0$.

An intriguing point  in the  evaluation of the coefficient $c_{\omega}$  in Ref. \cite{surowka} is
that it is mostly given in  thermodynamic terms. The only field-theoretic input is the chiral
anomaly in the presence of both electric and magnetic fields. A  puzzling observation is that the
coefficient $c_{\omega}$ is related to the coefficient in front of the  chiral anomaly  although the
chiral separation (\ref{effect}) persists in the absence of any electromagnetic fields. Some
progress in understanding relation of Eq. (\ref{effect}) to the field theory was made in Ref.
\cite{zahed} where it was observed that the chirality separation, similar to (\ref{effect}) arises
also in the flavor-related superfluidity (at zero temperature). The
 derivation \cite{zahed} is based  on the construction \cite{witten}
of the conserved currents corresponding to Wess-Zumino effective  action in the Goldstone phase.

 We address the issue of the chiral effects in the hydrodynamic approximation
 within an effective theory. The interaction in this effective theory is determined
 by the chemical potentials, $\mu$ and $\mu_{5}$ and the perturbation theory looks
 as an expansion in them. The first term in the expansion
  reproduces the leading contribution to $c_{\omega}$. Further terms appear
  to depend on details of the
 infrared regularization.

\section{Effective field theory.}

Our set up is as follows. There exists a fundamental Lagrangian constructed on massless left-handed,
$\psi_L$ and right-handed, $\psi_R$ fermions. The fundamental interaction is assumed to conserve
chirality and be non-anomalous, while in presence of external electromagnetic field $A_{\mu}$ there
is chiral anomaly. As an example one could think of two flavors of massless quarks interacting with
gluons in the standard flavor-blind way. This interaction is assumed to be responsible for formation
of dense quark-gluon matter state, whose relevant effective long-distance description is given in
terms of relativistic hydrodynamics. We choose microscopic currents as:
\begin{equation}\label{toy}
j^{i,\mu}~=~\bar{\psi}{\hat\tau}^i \gamma^{\mu} \psi~, \;\; j_5^{i,\mu}~=~ ~\bar{\psi}{\hat\tau}^i
\gamma^\mu\gamma_5 \psi.
\end{equation}
where the quark field $\psi = (u,d)$, ${\hat\tau}^0 = \hat{1}$ and ${\hat\tau}^{1,2,3}$ are the
standard Pauli matrices in flavor space.
  Classically these currents are conserved: $\partial_\mu j^\mu = \partial_\mu j_5^\mu = 0$.

To imitate the effects of the medium one introduces chemical potentials, conjugated to the
corresponding conserved charges:
 \begin{equation}
\delta H = \mu^i Q^i + \mu_5^i Q_5^i \label{ui}
\end{equation}
where $Q^i=\int d^3 x \psi^\dagger {\hat\tau}^i \psi$ and $Q_5^i=\int d^3 x \psi^\dagger \gamma_5
{\hat\tau}^i \psi$. At Lagrangian level (\ref{ui}) corresponds to:
  \begin{equation}\label{toy1}
 \delta L ~=~ \bar{\psi}{\hat\mu} \gamma^0 \psi ~+~ \bar{\psi}{\hat\mu}_5 \gamma^0 \gamma_5 \psi +
i \bar\psi {\hat\epsilon}(\mu,\mu_5,p_0)\> \psi .
 \end{equation}
where we have introduced the notation $\hat\mu = \mu^i {\hat\tau}^i $, ${\hat\mu}_5 = \mu_5^i
{\hat\tau}^i $. It is worth mentioning that the condition for the axial current to be non-anomalous
with respect to the strong interactions dictates its isovector nature, i.e. one must have $\mu_5^0 =
0$. There is no such restriction for the vector current which may contain singlet component. The
fact of currents conservation allows to introduce self-consistently the chemical potentials.

As is well known, the chemical potential $\mu$ conjugated to the charge $Q$ in quantum field theory
brings two aspects absent at zero density: first, it shifts energy levels of the system $p_0 \to p_0
- \mu$ (since all the levels below $\mu$ are occupied) and second, $i\epsilon$ prescription for
propagator poles changes to $\mu$ and $p$-dependent one (see, e.g. \cite{chodos}) and in this sense
the theory becomes non-local in coordinate space. It is worth stressing that both effects correspond
to a change of the vacuum state. However, as we will argue below, from effective theory point of
view they play rather different roles.

So far the discussion proceeded in terms of fundamental microscopic fields. We can now address the
most important issue of transition to effective theory. First, one is to introduce "physically
microscopic volume" of the medium, $V_y$, centered around the point $y$, in which infrared,
long-distance fields of effective theory may be regarded as being uniform. Performing Lorentz boost
to the rest frame one gets:
\begin{equation}\label{effective}
S = \int_{V_y} d^4x \;{\bar\psi} \gamma_\mu \left( i\partial^\mu + (\hat\mu + {\hat\mu}_5 \gamma_5)
u^\mu + {\hat{q}}A^\mu \right)\psi+ i\int_{V_y} d^4x\;{\bar\psi}{\hat\epsilon}(\mu,\mu_5,p,u)\psi + S_{int} \label{so}
\end{equation}
where $u^\mu$ is the four-velocity of a given element of the fluid in the center-of-energy rest
frame. As an infrared variable, it is supposed to be $x$-independent. We have also introduced
coupling to the external electromagnetic field $A^\mu$ with the charge matrix $\hat{q} =
\mbox{diag}(q_u, q_d)$. This field is also treated as long-distance non-dynamical one. The term
$S_{int}$ stays for strong interaction part of the total action whose exact form is not relevant
here.

To proceed one has to introduce summation over $V_y$ (i.e. integration over effective theory
coordinates $y$) and integrate out microscopic fields: quarks and gluons. However for our purposes
we need not to know the final result of this complex procedure. Instead, we notice that the effective
currents anomalous (non)conservation cannot be affected by $\mu$ dependence of $i\epsilon$-term in
(\ref{so}). The latter is an infrared effect vanishing at $\mu = 0$ and corresponding to vacuum
state change. At large momenta one always comes back to the standard $i\epsilon$ prescription with
constant $\epsilon$. This is completely analogous to the results on independence of chiral anomaly
on temperature/density found in \cite{ti}. Speaking more technically, coupling of the effective
scalar mass field $m-i\epsilon$ to fermions is non-anomalous regardless of its dynamics (encoded in
$\mu$-, $p$- and $u$-dependence of its complex $i\epsilon$ part in our case), even if the real part of the mass vanishes. Thus
the $\mu$, $\mu_5$ dependence of the anomaly can come only from the ${\bar\psi} \gamma_\mu (\hat\mu
+ {\hat\mu}_5 \gamma_5) u^\mu \psi$ term in the effective action. This term, on the other hand, can
be dealt with on exactly the same footing as the standard gauge field term ${\bar\psi} \gamma_\mu
{\hat{q}}A^\mu  \psi$.  For finding the anomaly it is convenient to use the Fujikawa-Vergeles method
\cite{bertlmann}. According to this method,  anomalies emerge due to non-invariance of the
path-integral measure under field transformations. For the sake of simplicity we consider below only diagonal components of the currents
 (i.e. $i=0,3$ components), however we continue to use invariant isospin notation. Consider the following transformation
\begin{equation}
\psi\rightarrow e^{i{\hat\alpha}\gamma_5+i{\hat\beta}}\psi
\end{equation}
One readily  finds \footnote{We could have defined anomaly in such a way that it does not contribute
to $\partial_{\mu}j^{\mu}_5$. In presence of both chemical potentials  $\mu$ and $\mu_5$ there is no
physical motivation for such a regularization.}:
\begin{eqnarray}\label{anomaly}
\partial_\mu j_5^{i,\mu}=-\frac{1}{4\pi^2}\epsilon_{\mu\nu\alpha\beta}{\mbox{Tr}}({\hat\tau}^i(
\partial^\mu
({\hat q}A^\nu+{\hat\mu} u^\nu)\partial^\alpha ({\hat q} A^\beta+ {\hat\mu}
u^\beta)+\partial^\mu({\hat\mu}_5 u^\nu)\partial^\alpha ({\hat\mu}_5 u^\beta)))
\end{eqnarray}
\begin{eqnarray}\label{anomaly3}
\partial_\mu j^{i,\mu}=-\frac{1}{2\pi^2}\epsilon_{\mu\nu\alpha\beta} {\mbox{Tr}} ({\hat\tau}^i
\partial^\mu
({\hat{q}} A^\nu+ {\hat\mu} u^\nu)\partial^\alpha ({\hat\mu}_5 u^\beta))~~,
\end{eqnarray}

In hydrodynamic approximation one has correspondence between microscopic currents $j^{i,\mu} =
\bar{\psi}{\hat\tau}^i \gamma^{\mu} \psi $, $ j_5^{i,\mu} = \bar{\psi}{\hat\tau}^i
\gamma^\mu\gamma_5 \psi$ and macroscopic effective ones $J^{i,\mu}$, $J_5^{i,\mu}$ carrying the same
conserved quantum numbers and given by (\ref{currents}). The collinear nature of the effective
currents is an important feature of non-superfluid hydrodynamics, where all charges densities
propagate with the same velocity $u^\mu$. Having in mind that the right hand sides of (\ref{anomaly})
and (\ref{anomaly3}) contain only effective long-distance fields one can replace divergencies of
microscopic currents $\partial j$ in the left hand sides by those of the effective currents
$\partial J$. In particular, the anomaly equation (\ref{anomaly}) can be rewritten as:
\begin{equation}\label{anomaly1}
\partial_\mu\left(n_5^i u^{\mu} + {\mbox{Tr}} {\hat\tau}^i \left( \frac{1}{2\pi^2}
({\hat\mu}^2+{\hat\mu}_5^2)\omega^\mu+
\frac{1}{2\pi^2}{\hat\mu} {\hat{q}} B^\mu \right)\right) \\
=-\frac{1}{4\pi^2}\epsilon_{\mu\nu\alpha\beta} {\mbox{Tr}}\left( {\hat\tau}^i {\hat q}^2 \partial^\mu
A^\nu\partial^\alpha A^\beta\right),
\end{equation}
where $B^\mu=\frac{1}{2}\epsilon^{\mu\nu\alpha\beta}u_\nu F_{\alpha\beta}$ is the magnetic field in the rest frame.
We will come back to discuss the implications of this equation later. Now, let us emphasize that in
the effective theory we have an expansion in $\mu,\mu_5$. The anomalous terms are exhausted by the
triangle graph, or by terms quadratic in $\mu,\mu_5$ kept explicit in
 (\ref{anomaly3}), (\ref{anomaly1}). Another observation is that to keep the coefficient of the term
 $\omega^\mu$
non-vanishing one must have nonzero singlet component in the vector current (otherwise ${\mbox{Tr}}
{\hat\tau}^i {\hat\mu}^2 = 0$). The result for the coefficient $c_{\omega}$ implied by
(\ref{anomaly1}) coincides with that of the paper \cite{surowka} in the lowest non-trivial order in
the chemical potentials. The comparison is not absolutely straightforward, though. The reason is
that we introduce two chemical potentials, $\mu,\mu_5$, and assume interactions to conserve parity
(so that $\mu_5$ is actually a pseudoscalar).   On the other hand, the simplest version considered
in most detail in \cite{surowka} deals with a single chiral current, with no parity conservation

It is worth emphasizing that there exist contributions of higher orders in $\mu,\mu_5$ to the
hydrodynamic currents. The higher-order terms, however,
 do not contribute  to the anomalous divergences evaluated within the effective theory
 and for this reason do not enter Eq (\ref{anomaly}).
Moreover, such terms depend, generally speaking, on details of the infrared cutoff and we make no
attempt to find them in closed form. For the sake of an estimate, consider the $\mu^3$ contribution
to the $\omega^{\mu}$ term. It is of the order:
\begin{equation}\label{divergent}
\delta c_{\omega}~\sim~ {\mu^3\over 2\pi^2}{1\over \epsilon_{IR}}~,
\end{equation}
where we keep the $(2\pi^2)^{-1}$ factor just to indicate that it is a one-loop correction and
$\epsilon_{IR}$ is an infrared cutoff in the energy/momentum integration.
 In the hydrodynamic limit, the following estimate for $\epsilon_{IR}$
 seems reasonable:
$$\epsilon_{IR}~\sim~(\epsilon+p)/n~~,$$
 where $\epsilon,p$ are the energy density and pressure respectively and $n$
 is the particle density. Moreover, the ratio $(\epsilon+p)/n$
 is known to play the
 role of the mass in the relativistic hydrodynamic.
 The effective-theory estimate  (\ref{divergent})
 reproduces then the structure of  the explicit contribution to $c_{\omega}$
 found in \cite{surowka}, $\delta c_{\omega}~\sim~\mu^3n/(\epsilon+p)$.

Our final remark in this section is that the expression (\ref{anomaly3}) for the vector current is
closely related to another chiral effect in the hydrodynamic approximation, that is the chiral
magnetic effect thoroughly discussed in the literature,
 for a review and references see \cite{kharzeev}.
In particular, in case $u^0=1, u^a=0$ and with one flavor we have
$$J_a~=~ ~- \frac{1}{2\pi^2}\>\mu_5 B_a \>,$$
where $a=1,2,3$ and $B_a=\frac{1}{2}\epsilon_{abc}F^{bc}$ are components of the external magnetic field, and reproduce well known
results \cite{kharzeev} for the chiral magnetic effect. In case of nontrivial $u^{\mu}$ Eq
(\ref{anomaly}) demonstrates existence of hydrodynamic corrections. For more details see
\cite{isachenkov}.

\section{Comparison with thermodynamic approach}

We can compare the results obtained within the effective theory (\ref{effective}) with the results
of Ref \cite{surowka} which are entirely based on thermodynamics (plus equations of state). The
results within the thermodynamic approach can be summarized in the following way:
\begin{eqnarray}
J^{\mu}_{ch}=n_{ch}u^\mu~+ ~\xi_{\omega}\omega^\mu~+~\xi_B B^\mu
\end{eqnarray}
where $J^{\mu}_{ch}$ is a chiral current associated with, say, left-handed fermions. Imposing the
positivity of the flow of the entropy one fixes the coefficients uniquely. In particular,
\begin{equation}\label{thermodynamics}
\xi_{\omega}~=~{\mu^2_{ch}\over 2\pi^2}\Big(1-\frac{2}{3}{\mu_{ch}n_{ch}\over \epsilon +p}\Big)~,
\end{equation}
where $n_{ch}$ and $\mu_{ch}$ are the  density of the chiral particles and the corresponding
chemical potential, respectively.

One can readily see that the $\mu^2_{ch}$ terms are the same as within the effective-theory approach
above. However, for higher order terms the predictions vary. According to the effective-theory
approach the $\mu^3$ terms are dependent on details of the infrared cut off and not fixed in this
sense. According to (\ref{thermodynamics}), on the other hand, the $\mu_{ch}^3$ terms are uniquely
determined.

To appreciate the meaning of this discrepancy one should have in mind that some further assumptions
were made in \cite{surowka}. In particular, the ideal-liquid equations of motion are used. More
generally, as is emphasized in \cite{surowka} the vorticity term $\omega^{\mu}$ appears within the
standard hydrodynamic approximation in higher orders in derivatives than the leading term, which is
$J^{\mu}_{ch}\approx n_{ch}u^{\mu}$. Keeping systematically all the terms of the next order in
derivatives (not only $\omega^{\mu}$) would modify equations of motion and destroy
(\ref{thermodynamics}).
 Moreover, as is noticed in \cite{dtson} there is a kind of   or gauge
 invariance in relativistic formulation of hydrodynamics which allows to redefine
 higher-derivatives terms in expansion of the energy-momentum tensor and of currents.

Thus, we believe that it is the extra assumptions made which allow to fix the $\mu^3$ terms within
the thermodynamic approach. Within the effective theory approach the $\mu^3$ terms are infrared
dependent. It is worth mentioning that the infrared dependence of higher order terms was found also
within a superfluid version of hydrodynamics considered in \cite{zahed} \footnote{Actually the Ref
\cite{zahed} does not state explicitly that the $\mu^3$ terms are not uniquely fixed. Rather, a
particular prescription for an infrared cut off is picked up.}.

\section{ Conservation laws in hydrodynamic approximation}

  To summarize, within the model considered we have two types of anomalies.
  First, there is the common chiral anomaly in the
  divergence of the  fundamental axial current,
  see Eq. (\ref{anomaly}). Second,
   there are anomalies presented in the effective
  theory alone.

  Let us first consider the case when the product
  ${\bf (E\cdot H)}=0$ for external fields and the fundamental
  currents are conserved. The origin of the anomalies in
  the effective theory is that the
  symmetry of the effective theory is not the same as of the
  fundamental interaction. The point is clearly illustrated by the
  toy model (\ref{toy}), (\ref{toy1}). The strong interaction
  of quarks and gluons is flavor blind. We choose however
  asymmetric initial conditions by fixing the average number
  of $u$-quarks but not of $d$-quarks. Introduction of the
  effective interaction (\ref{toy1})
  transfers this flavor-asymmetry of the initial condition into
  the flavor-dependent effective
  interaction. Hence, anomalies emerge in the language
  of the effective theory.

  The first example of this type was seemingly given in Ref. \cite{witten}.
  In that case, effective Lagrangian for the interaction of Goldstone bosons $\pi^a$
  is constructed in terms of the matrix
  $$U=\exp \Big( {2i\over F_{\pi}}\lambda^a\pi^a \Big)~,~~~
  U_{\mu}^R~\equiv~U^{-1}\partial_{\mu}U~.$$
 The conservation condition is then a sum of two terms, naive and anomaly-like ones:
 \begin{equation}\label{witten}
 -\partial_{\mu}{F_{\pi}^2U_{\mu}^R}+(i/2)\epsilon_{\mu\nu\alpha\beta}
 U^R_{\mu}U^R_{\nu}U_{\alpha}^RU_{\beta}^R~=~0.
\end{equation}
 As is noted first in Ref. \cite{zahed} the analogy
 between (\ref{witten}) and the hydrodynamic models
 we are considering might  be much more direct than it appears
 at first sight.
 We come back to discuss this point later.

 Note that the effective theory fixes all the anomalous terms in the
 left-hand side of the condition
 (\ref{anomaly1}) in the absence of the genuine chiral anomaly
 which arises in case of ${\bf (E\cdot H)}\neq 0$. This is again in
 analogy with the case considered in Ref. \cite{witten} and in contrast
 with the thermodynamic derivation of Ref. \cite{surowka} where the
 genuine chiral anomaly is a necessary input to fix the vorticity
 coefficient $c_{\omega}$.
 The simplicity of the Eq. (\ref{anomaly1}) is somewhat deceptive,
 however, since the current itself, generally speaking, contains
 divergence-free terms dependent on the infrared cut off.

 Turn now to the case of ${\bf (E\cdot H)}\neq 0$.
 Then there is an immediate problem that the axial current $j_{\mu 5}$
 is no longer conserved and the introduction of the chemical potential, see
   (\ref{effective}), conjugated to a non-conserved charge
 is not consistent, for recent discussion see \cite{rubakov} and references therein.
As is known since long, see in particular \cite{gellmann} one can introduce  a new conserved axial
current also in the presence of external fields with ${\bf (E\cdot H)}\neq 0$:
\begin{equation}
\partial_{\mu}\tilde{J}_5^\mu~=~0~,
\end{equation}
where
\begin{equation}\label{redefinition}
\tilde{J}_5^\alpha~=~n_5 u^{\alpha}~+~{1\over 2\pi^2}(\mu^2+\mu^2_5)\omega^{\alpha}+{\mu\over 2\pi^2}
B^{\alpha}~-~K^{\alpha}~~,
\end{equation}
where $B^{\mu}\equiv \tilde{F}^{\mu\nu}u_{\nu}$. Although the current $K_{\mu}$ is not gauge
invariant, the corresponding charge, $\int d^3x K_0$ is gauge invariant. However, we are considering
the approximation of external electromagnetic fields and the value of ${\bf (E\cdot H)}$ is not
dynamical. In this sense the chiral charge of the medium is changed under influence of external
fields, independent of possible redefinition (\ref{redefinition}). Therefore Eq (\ref{anomaly1}) is
rather to be understood as an approximate, in the sense that we neglect the change of the axial
charge due to the ${\bf (E\cdot H)}\neq 0$ (the change of the axial charge of the medium is
proportional to time for constant ${\bf E}$ and ${\bf H}$).

Turn now to a more delicate point, that is how the 't Hooft matching condition \cite{thooft} is
realized in the hydrodynamic approximation. Let us begin with commonly accepted points. At vanishing
temperatures, the alternatives are that  there exist either massless, not confined quarks or
massless pseudoscalars. Clearly, in QCD the Goldstone mode is realized at $T=0$. At small finite
temperatures there are well defined corrections to the pseudoscalar  decays constants
\cite{pisarski}. Finally, at the temperature of deconfining phase transition $T=T_c$ pions become
massive and the matching involves massless, not confined quarks.

 Now we are considering hydrodynamic approximation
 which assumes averaging over distances $\Delta x$ larger than
 the free path of the quarks:
 $$\Delta x~\gg l_{free~path}$$
 As a result of the interaction liquid is formed and
 one considers hydrodynamics, that is a classical approximation.
 The central point is then that there are no fermionic classical fields.
 In other words , in the hydrodynamic approximation there can be no
 fermionic (massless) excitations which would trivially saturate
 the 't Hooft matching condition.
 Thus, we are led to conclude that in the hydrodynamic approximation
 the 't Hooft consistency condition is to be satisfied on the bosonic massless
 modes \footnote{In reality, even the light $u$- and $d$-quarks might be too heavy
 to be considered  massless on the distances $\Delta x$ relevant to the hydrodynamics.
 }.

To reiterate, for fermions interacting only with external magnetic field the chiral anomaly is
realized through chiral zero modes, which are nothing else but particular Landau levels, for details
see \cite{ninomiya,kharzeev}. However, now we consider the case when the fermions interact among
themselves and form a  liquid as a result of this interaction. This approximation corresponds to
\begin{equation}\label{landau}
l_{free~path}~<~R_{Landau}~,
\end{equation}
where $l_{free~ path}$ is the free-path length in the liquid and $R_{Landau}$ is the radius of the
would-be zero-energy Landau level. Note also that the hydrodynamic approximation assumes coarse
graining at scale larger than $l_{free~path}$. Then the fermions lose coherence and Landau levels do
not exist as solutions any longer. Therefore in the hydrodynamic approximation the matching of the
chiral anomaly to physical fermionic excitations is questionable.

Thus, matching of the quark anomaly to massless pseudoscalar degrees of freedom seems to be a viable
alternative. In that case one would introduce an analog, $\tilde{U}_{\mu}^R$, to the matrix
$U_{\mu}^R$ above and construct an effective current in terms of a new field $\pi^a_{hydro}$ which
might be called `hydrodynamic shadow' of the pion. As is argued in Ref. \cite{zahed} the two terms
in the current (\ref{witten}) can be interpreted in the hydrodynamic approximation as naive current
$n_5u_{\mu}$ and the vorticity term $c_{\omega} \omega_{\mu}$.

In our case, introduction of a specifically hydrodynamic massless excitations $\pi_{hydro}^a$ would
result, generally speaking, in a two-component liquid:
\begin{equation}
j^\mu_5~\sim~n_5u^{\mu}~+~\tilde{n}_5v^{\mu}+...~~,
\end{equation}
where $u_{\mu}, v_{\mu}$ are two independent 4-velocities. Relativistic version of the hydrodynamics
of  two-component liquids has been intensely discussed recently, see in particular \cite{herzog}.
Generically, existence of massless bosonic excitations results in superfluidity.

\section{Conclusion.}

 In this note we have attempted to evaluate the newly discovered hydrodynamic chiral effects
within the effective-theory approach. The perturbation theory within this approach is an expansion
in chemical potentials, $\mu,\mu_5$. The effective theory is anomalous. The chiral anomaly fixes
uniquely the $\mu^2$ terms. Moreover, the anomaly is exhausted, as usual, by its lowest non-trivial
order. The $\mu^2$ terms generated by the anomaly
  coincide
with the $\mu^2$ terms found earlier within a pure thermodynamic approach \cite{surowka}.
  Higher-order terms turn to be infrared dependent within the effective-theory approach.

  Our interest in the problems considered stems not so much from phenomenological
  applications but rather from the fact that hydrodynamics appears to provide a novel example
  of realization of chiral symmetries. Namely, the symmetry which is not anomalous in the
  fundamental theory turns to be anomalous within the hydrodynamic approximation.
   One of the mechanisms of generating anomalous effective theories is the
   introduction of  chemical-potential terms. Originally, chemical potentials
   reflect initial conditions, in terms of conserved charges. The symmetry
   of the initial conditions does not necessarily coincide with symmetry
   of the fundamental interactions. Also, we argued that the 't Hooft matching
   condition in the hydrodynamics favors massless boson excitations. Thus, even
   at temperatures above the deconfining phase transition there could exist
   specific hydrodynamic excitations $\pi^a_{hydro}$ with quantum numbers of ordinary pions.

We are thankful to  A.S. Gorsky, F.V. Gubarev and M.V. Isachenkov  for discussions.
The work of A.V. Sadofyev and of V.I. Zakharov  was partially supported by RFBR grant no. 10-02-01483.
The work of Sadofyev was also partially supported by DAAD Leonhard-Euler-Stipendium 2010-2011.

\end{document}